\documentclass[12pt]{article}

\usepackage[dvips]{graphicx}
\renewcommand{\baselinestretch}{1.75}
\newcommand \be{\begin{equation}}
\newcommand \ee{\end{equation}}
\newcommand \ba{\begin{eqnarray}}
\newcommand \ea{\end{eqnarray}}
\begin{document}

\def\today{\ifcase\month\or
 January\or February\or March\or April\or May\or June\or
 July\or August\or September\or October\or November\or
December\fi
 \space\number\day, \number\year}
%

\hfil PostScript file created: \today{}; \ time \the\time \
minutes
\vskip .15in

\centerline {STATISTICAL EARTHQUAKE FOCAL MECHANISM FORECASTS}

\vskip .15in
\begin{center}
{Yan Y. Kagan and David D. Jackson}
\end{center}
\centerline {Department of Earth and Space Sciences,
University of California,}
\centerline {Los Angeles, California 90095-1567, USA;}
\centerline {Emails: {\tt kagan@moho.ess.ucla.edu,
david.d.jackson@ucla.edu}}

\vskip 0.12 truein

Accepted date. Received date; in original form date

\vskip .15in
\noindent
{\bf Short running title}:
{\sc
Focal Mechanism Forecasts
}

\vskip .15in

Corresponding author contact details:

Office address:  
Dr. Y. Y. Kagan, Department Earth and Space Sciences (ESS), 
Geology Bldg, 595 Charles E. Young Dr.,
University of California Los Angeles (UCLA), 
Los Angeles, CA 90095-1567, USA 

Office phone  310-206-5611 

FAX:          310-825-2779 

E-mail: kagan@moho.ess.ucla.edu 

\vskip 0.25in

SECTION: SEISMOLOGY

\newpage

\noindent
{\bf SUMMARY}
\hfil\break
Forecasts of the focal mechanisms of future earthquakes are
important for seismic hazard estimates and Coulomb stress and
other models of earthquake occurrence. 
Here we report on a high-resolution global forecast of
earthquake rate density as a function of location, magnitude,
and focal mechanism. 
In previous publications we reported forecasts of 0.5 degree
spatial resolution, covering the latitude range from $-$75 to
+75 degrees, based on the Global Central Moment Tensor
earthquake catalog. 
In the new forecasts we've improved the spatial resolution to
0.1 degree and the latitude range from pole to pole. 
Our focal mechanism estimates require distance-weighted
combinations of observed focal mechanisms within 1000 km of
each grid point. 
Simultaneously we calculate an average rotation angle between
the forecasted mechanism and all the surrounding mechanisms,
using the method of Kagan \& Jackson proposed in 1994.
This average angle reveals the level of tectonic complexity of
a region and indicates the accuracy of the prediction. 
The procedure becomes problematical where longitude lines are
not approximately parallel, and where earthquakes are so
sparse that an adequate sample spans very large distances. 
North or south of 75 degrees, the azimuths of points 1000 km
away may vary by about 35 degrees. 
We solved this problem by calculating focal mechanisms on a
plane tangent to the earth's surface at each forecast point,
correcting for the rotation of the longitude lines at the
locations of earthquakes included in the averaging. 
The corrections are negligible between $-$30 and +30
degrees latitude, but outside that band uncorrected rotations
can be significantly off.
Improved forecasts at 0.5 and 0.1 degree resolution are posted
at http://eq.ess.ucla.edu/~kagan/glob\_gcmt\_index.html.

\vskip 0.05in
\noindent
{\bf Key words}:
\vskip .05in
Probabilistic forecasting;
Earthquake interaction, forecasting, and prediction;
Seismicity and tectonics;
Theoretical seismology;
Statistical seismology;
Dynamics: seismotectonics.

\newpage

\section{Introduction}
\label{intro}
This paper addresses two problems: forecasting earthquake
focal mechanisms and evaluating forecast skill. 
Properties of earthquake focal mechanisms and methods for
their determination are considered by Snoke (2003) and
Gasperini \& Vannucci (2003). 

The focal mechanism forecast method was originally developed
by Kagan \& Jackson (1994). 
Kagan \& Jackson (2000, 2011) applied this method to regional
and global seismicity forecasts inside the latitude band $[
75^\circ~S - 75^\circ~N ]$. 
In the present forecast, the weighted sum of normalized
seismic moment tensors within 1000~km radius is calculated and
the T- and P-axes for the predicted focal mechanism are
evaluated by calculating summed tensor eigenvectors. 
We also calculate an average rotation angle between
the forecasted mechanism and all the surrounding mechanisms. 
This average angle shows tectonic complexity of a region and
indicates the accuracy of the prediction. 

Recent interest by CSEP (Collaboratory for the Study of
Earthquake Predictability) and GEM (Global Earthquake Model)
has motivated some improvements, particularly to extend the
previous forecast to polar and near-polar regions. 
For more information on CSEP see Eberhard {\it et al.}\ 
(2012), Zechar {\it et al.}\ (2013 and references therein). 
The GEM project is briefly described by Storchak {\it et al.}\
(2013). 

The major difficulty in extending the forecast beyond the $
[75^\circ~S - 75^\circ~N ]$ latitude band is convergence of 
longitude lines in polar areas.
To take it into account we need to account for bearing
(azimuth) difference within the 1000~km circle that we used
for averaging seismic moment tensors. 
We consider the bearing correction and apply it in averaging 
seismic moment tensors.
In most situations a forecast point where we calculate an
average focal mechanism is surrounded by earthquakes, so a
bias should not be strong due to the difference effect
cancellation. 
We show that such a correction improves the forecast in 
near-polar areas.

The skill of a focal mechanism forecast can be measured by 
studying the distribution of orientation difference between 
predicted and measured focal mechanisms.
We investigate this difference by measuring a 3-D rotation
angle (Kagan 1991) between those earthquake sources.
In particular, a predicted focal mechanism is evaluated on the
basis of the GCMT catalog for 1977-2007, and we measure the
angle for earthquake measurements during 2008-2012. 
Thus, this is a pseudo-prospective test, and we plan to carry 
out real-time prospective test.

In addition to constructing a focal mechanism forecast and
studying its performance, we need to evaluate the complexity
of the forecasted moment tensor. 
For this purpose we measure the non-double-couple or CLVD
component of the tensor.
The $\Gamma$-index (Kagan \& Knopoff 1985a) is the best method
to accomplish this. 
A non-zero index indicates that earthquake focal mechanisms
around the forecast point have different orientations. 
We compute the index and analyze its correlation with the
rotation angle for all the predicted points. 
Thus deformation complexity displays itself in the average
rotation angle and in the index. 

As the final result of these investigations we construct the 
whole Earth forecast in two formats: medium- and
high-resolution.
Both new $0.5 \times 0.5^\circ$ and $0.1 \times 0.1^\circ$
forecasts are posted at
http://eq.ess.ucla.edu/$\sim$kagan/glob\_gcmt\_index.html.

Input catalog data are described in Section~\ref{ctlg}. 
Section~\ref{fearth_long} considers methods used in creating 
focal mechanism forecasts.
In Section~\ref{bearing} we discuss a method for computation
of bearing difference between two points on a sphere. 
Section~\ref{forecast} describes methods for measuring the 
focal mechanism forecast performance or skill.
Section~\ref{comp} is dedicated to the statistical analysis of 
3-D rotation angle and the $\Gamma$-index for characterization 
of focal mechanism complexity.
Section~\ref{disc} summarizes our results and suggests 
techniques for improving focal mechanism forecasts.

\section{Data}
\label{ctlg}

\begin{figure}
\begin{center}
\includegraphics[width=0.60\textwidth,angle=-90]{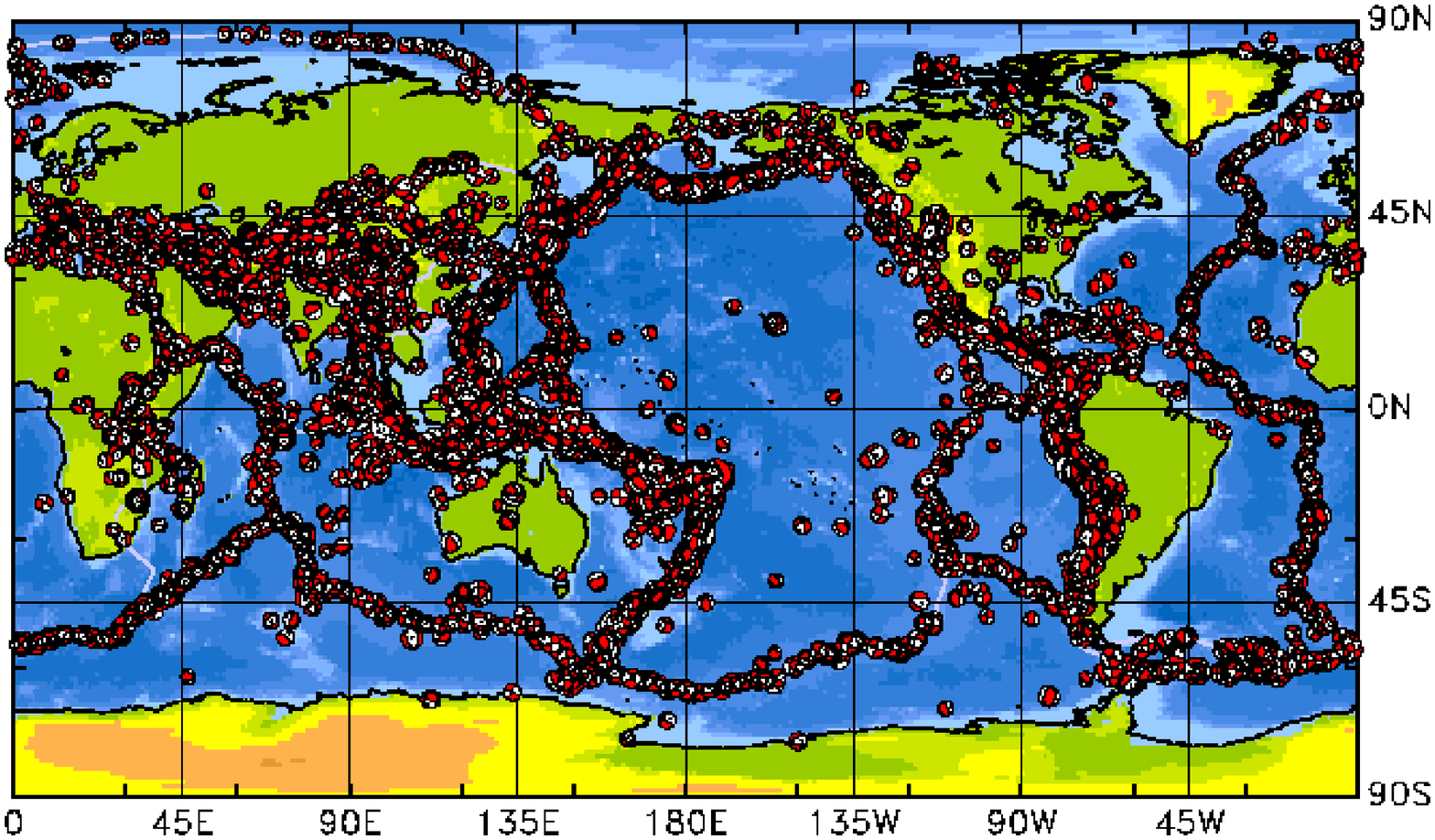}
\caption{Location of shallow (depth 0--70~km) earthquakes }
\label{fig01}
\end{center}
\vskip -.5cm
in the Global Centroid Moment Tensor (GCMT) catalog,
1976--2012.
Earthquake focal mechanisms are shown by stereographic
projecting of the lower focal hemisphere (Aki \& Richards
2002).
The size of a symbol is proportional to earthquake magnitude.
(Courtesy of G\"oran Ekstr\"om and the GCMT project).
\end{figure}

In Fig.~\ref{fig01}, we display a map of earthquake
centroids in the global CMT (Centroid-Moment-Tensor)
catalog (Ekstr\"om {\it et al.}\ 2012, and its
references).
The earthquakes in the catalog are mostly concentrated at
tectonic plate boundaries.
Each earthquake is characterized by a centroid moment tensor
solution.

The present catalog contains more than 38,000 earthquake
entries for the period 1976/1/1 to 2012/12/31.
Earthquake size is characterized by a scalar seismic moment
$M$.
Earthquake moment magnitude $m_W$ is related to the scalar
seismic moment $M$ via (Kanamori 1977)
\be
m_W = \frac{2}{3} \log_{10} M - C \, ,
\label{Eq0}
\ee
where seismic moment $M$ is measured in Newton-m, and
$C$ is usually taken to be between 6.0 and 6.1.
Below we use $C=6.0$ (Hanks 1992).
We consider the full catalog to be complete above our lower
threshold magnitude $m_t=5.8$ (Kagan 2003). 

\begin{figure}
\begin{center}
\includegraphics[width=0.65\textwidth]{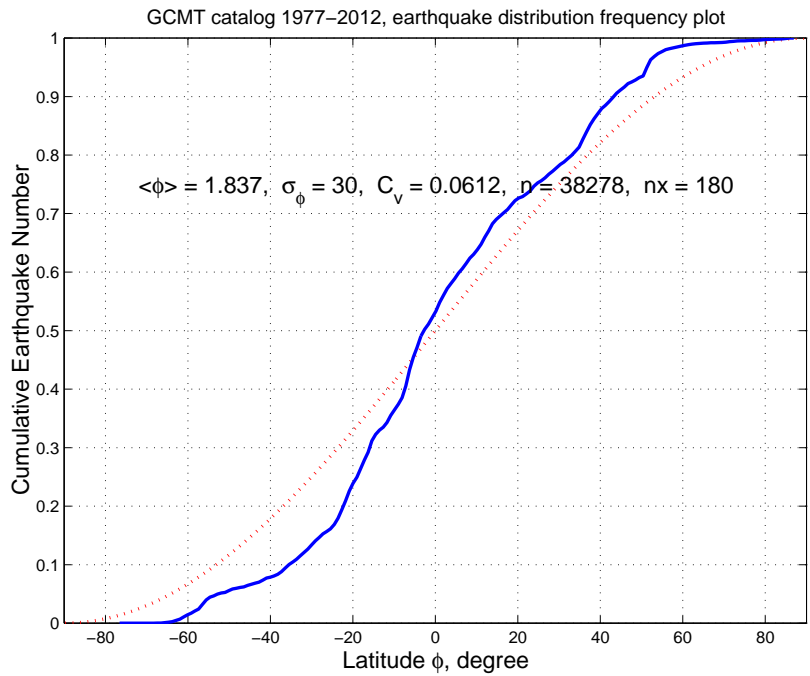}
\caption{GCMT catalog, 1977--2012.
} 
\label{fig03}
\end{center}
\vskip -.5cm
Blue solid line is cumulative latitudinal distribution of
earthquakes. 
Red dotted line corresponds to spherical equal-area earthquake
distribution.
\end{figure}

Fig.~\ref{fig03} displays the distribution of earthquakes in
the 1977-2012 GCMT catalog by latitude.
Earthquakes are concentrated more in equatorial areas compared
to an equal-area distribution, and there are more events in the
northern hemisphere (compare Fig.~\ref{fig01}).
There are few earthquakes (around 5\%) below 50$^\circ$S, but
in the northern polar area the number is greater, close to
8\%.
As we will see later, these near polar focal mechanisms would
need a special correction to calculate the forecasted
mechanism.

\section{Long-term focal mechanism estimates}
\label{fearth_long}

Kagan \& Jackson (1994) present long-term earthquake forecasts
in several regions using the GCMT catalog.
They use a spatial smoothing fixed kernel proportional to
inverse epicentroid distance $1/r$, except that it was
truncated at short and very long distances (see their Fig.~4).
Jackson \& Kagan (1999) and Kagan \& Jackson (2000; 2011) use
a different power-law kernel 
\be
f (r) \ = \ {1 \over \pi} \times {1 \over { {r^2 + r_s^2} } }
\, ,
\label{eq1}
\ee
where $r$ is epicentroid distance, $r_s$ is the scale
parameter, taken here $r_s = 2.5$~km, and we select $r \le
1000$~km to make the kernel function integrable. 
Kagan \& Jackson (2012) consider several kernel functions 
including the adaptive Fisher kernel specifically conformed to
the spherical surface. 
Our procedure (Kagan \& Jackson 1994, 2000, 2011) allows us to
optimize the parameters by choosing those $r_s$ values which
best predict the second part of a catalog (test or validation
period), using a maximum likelihood criterion, from the first
part (training or learning period). 

\begin{figure}
\begin{center}
\includegraphics[width=0.65\textwidth]{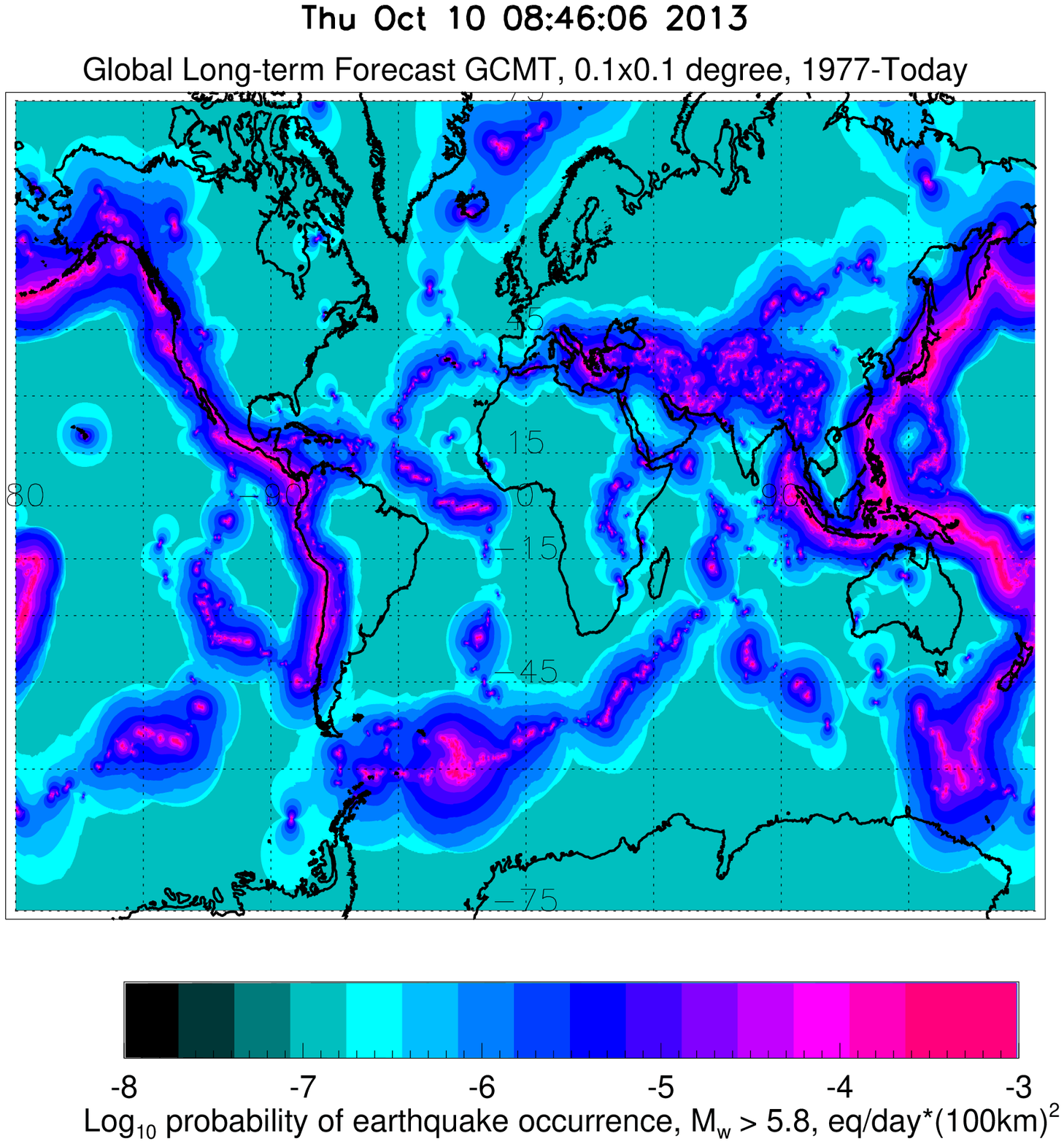}
\caption{Global earthquake long-term potential based on
}
\label{fig04}
\end{center}
\vskip -.5cm
smoothed seismicity, latitude range $[ 75^\circ S-75^\circ N
]$ at $0.1 \times 0.1^\circ$ spatial resolution. 
Earthquakes from the GCMT catalog since 1977 are used.
Earthquake occurrence is modelled by a time-independent
(Poisson) process.
Colors show the long-term probability of earthquake
occurrence.
\end{figure}

The kernel (\ref{eq1}) is elongated along the fault-plane,
which is estimated from the available focal mechanism
solutions. 
To accomplish this we multiply the kernel by an orientation
function $D (\varphi)$ depending on the angle $\varphi$
between the assumed fault plane of an earthquake and the
direction to a map point
\be
D (\varphi) \ = \ 1 \ + \ \Theta \times \cos^2(\varphi) \, .
\label{eq2}
\ee
The parameter $\Theta$ above controls the degree of azimuthal
concentration (Kagan \& Jackson 1994, their Fig.~2), we take
$\Theta = 25$. 
In smoothing we weigh each earthquake according to its moment
magnitude
\be
w \ = \ m / m_t \, ,
\label{Eq_inf-21}
\ee
where $m_t$ is a magnitude threshold, $m_t = 5.8$.
Fig.~\ref{fig04} demonstrates such a forecast of the global
long-term earthquake rate density for magnitudes of 5.8 and
above, based on the GCMT catalog.

We forecast the focal mechanism of a predicted earthquake
following Kostrov's (1974) suggestion: first we predict the
focal mechanism of an earthquake by summing up the past
events, with a weighting as given above:
\be
M_{pq} \ = \ {\sum^n_{i=1}} \, ( M_{pq})_i \, f (r_i) \,
D(\varphi_i) \, w_i \, , 
\label{eq3}
\ee
where $(M_{pq})_i$ is the normalized seismic moment
tensor of the $i$-th earthquake in the catalog. 
Then we calculate the eigenvectors of the sum $ M_{pq}$ and
assign the eigenvector corresponding to the largest eigenvalue
as the $T$-axis, and the eigenvector corresponding to the
smallest eigenvalue as the $P$-axis of a forecasted event. 

Thus, although $M_{pq}$ in general is not a double-couple, we
take the normalized double-couple ($DC$) component of the
tensor as the forecasted mechanism. 
The angular density function of $M_{pq}$ is a function of
the $n$ minimum 3-D rotation angles $0^\circ \le \Phi_1
\le 120^\circ $ (Kagan 1991) necessary to transform each of
the observed focal mechanisms into the predicted one
(\ref{eq3}). 
The weighted average rotation angle $\Phi_1$ shows the degree
of tectonic complexity at this point. 
Our forecast tables show the plunge and azimuth of the $T$-
and $P$-axes as well as the rotation angle $ \, \Phi_1 \, $
(see
http://eq.ess.ucla.edu/$\sim$kagan/glob\_gcmt\_index.html). 
An example of the forecast table is shown by Kagan \& Jackson
(2000, their table~1).

\section{Bearing difference correction }
\label{bearing}

The bearing (or azimuth) of a tangent vector through a point on
a sphere (Richardus \& Adler 1972) is the angle between that
vector and the longitude line through the point. 
Near the equator, the bearing $\beta_{12}$ from point 1 to
point 2 is different by about 180$^\circ$ from the bearing
$\beta_{21}$ from point 2 to point 1, because the longitude
lines at the two points are nearly parallel. 
However, near the poles the longitude lines could be far from
parallel, and we need to correct for the bearing difference
$\Delta \beta$. 

To compute the bearing difference at two points on the Earth
surface, we calculate 
\be
S_1 \ = \ \cos(\phi_2) \times \sin(\psi_2-\psi_1)
\, ,
\label{eq14}
\ee
where $\phi_i$ is latitude and $\psi_i$ is longitude of the 
points,
\be
C_1 \ = \ \cos(\phi_1) \times \sin(\phi_2) \ - \
\sin(\phi_1) \times \cos(\phi_2) \times \cos(\psi_2-\psi_1)
\, ,
\label{eq15}
\ee
\be
\beta_{12} \ = \ {\rm mod} \, [ \, \arctan2 \, (S_1, \,
C_1)+360^\circ, \, 360^\circ \, ]
\, ,
\label{eq18}
\ee
where 
$\arctan2 \, (S_1, \, C_1)$ returns angle values in the range
$[ \, -180^\circ ... +180^\circ \, ]$, and 
${\rm mod} \, (x_1, x_2) $ is the remainder when $x_1$ is 
divided by $x_2$, i.e., $x_1$ modulo $x_2$.
Similarly
\be
S_2 \ = \ \cos(\phi_1) \times \sin(\psi_1-\psi_2)
\, ,
\label{eq16}
\ee
\be
C_2 \ = \ \cos(\phi_2) \times \sin(\phi_1) \ - \
\sin(\phi_2) \times \cos(\phi_1) \times \cos(\psi_1-\psi_2)
\, ,
\label{eq17}
\ee
\be
\beta_{21} \ = \ {\rm mod} \, [ \, \arctan2 \, (S_2, \,
C_2)+180^\circ, \, 360^\circ \, ]
\, ,
\label{eq19}
\ee
and
\be
\Delta \beta \ = \ \beta_{21} \, - \, \beta_{12}
\, .
\label{eq20}
\ee
To calculate the focal mechanism parameters in the tangent
plane we add $\Delta \beta$ to azimuths of $T$- and $P$-axes,
and recompute the seismic moment tensor. 

Fig.~\ref{fig04a} displays forecasted focal mechanisms, 
similarly as was done by Kagan \& Jackson (1994, their 
Figs.~6a,b).
To avoid the figure congestion the mechanisms are shown at 
$5^\circ \times 5^\circ $ grid, but they are calculated at 
$0.5^\circ \times 0.5^\circ$ or $0.1^\circ \times 0.1^\circ$
spatial resolution. 
We exclude from the display the areas where no earthquake was 
registered within 1000~km distance.
These regions are shown by the greenish-gray color in
Fig.~\ref{fig04}. 
In these areas our forecast tables specify a ``default" focal
mechanism output 
\be
T = 0^\circ, \, 180^\circ; \ P = 90^\circ, \, 90^\circ; \quad
{\rm and} \quad \Phi_1 = 0.0; 
\quad \Gamma = 0.0 
\, ,
\label{eq21}
\ee
where the first item in $T$- and $P$-axes is a plunge and 
the second one is an azimuth (Aki \& Richards 2002, Figs.~4.13
and 4.20). 

Comparing these predicted mechanisms with their actual 
distribution in Fig.~\ref{fig01} demonstrates that our
forecast reasonably reproduces earthquake sources properties. 
The forecast advantage is that the prediction accuracy is 
evaluated.

\section{Focal mechanism forecast skill }
\label{forecast}

To evaluate the skill of the focal mechanism forecast we
subdivide the GCMT catalog into two parts: 1977-2007 and
2008-2012.
The first part was used to calculate the expected focal
mechanism at all the epicentroids of 1977-2012 earthquakes,
then we estimate how the observed mechanisms of 2008-2012
period differ from the prediction. 
To accomplish this we measure the minimum 3-D rotation angle
$\Phi_2$ between these double-couples (Kagan 1991). 

Fig.~\ref{fig06} shows the cumulative distribution of the
$\Phi_1$ angle which is the average rotation angle between the
weighted focal mechanisms (\ref{eq3}) and mechanisms of the
1977-2007 earthquakes in a 1000~km circle surrounding this
forecasted event. 
For about 90\% of the forecasts the average angle $\Phi_1$ is
less than 45$^\circ$. 
The average angle ($< \Phi_1 >$) and its standard deviation
($\sigma_\Phi$) are also shown. 

For comparison in Fig.~\ref{fig06} we display two theoretical
angle distributions: the rotational Cauchy distribution (Kagan
2000) and the purely random rotation of a $DC$ source (Kagan
1991). 
Kagan (2000) argues that in the presence of random defects in
solids, rotation angles should follow the Cauchy law. 
The Cauchy distribution has only one parameter ($\kappa$), and
it approximates the $\Phi_1$ curve reasonably well up to about
20$^\circ$.
If the observed $\Phi_1$ curve were close to the random
rotation distribution for the $DC$ source, this would mean
that there is no useful information in the forecast. 
The former curve is significantly different from the latter 
one, demonstrating good forecast skill. 

In Fig.~\ref{fig07} the distribution for the $\Phi_2$ angle
is displayed; this is the angle between the predicted
mechanism (\ref{eq3}) and the $DC$ mechanism of the observed
events in the 2008-2012 period. 
Angle $\Phi_1$ in Fig.~\ref{fig06} is usually smaller than the
observed angle $\Phi_2$, because the former angle is an
average of many disorientation angles, whereas the latter
angle corresponds to just one observation. 
The distribution difference of two angles can be seen in their
averages ($ < \Phi > $) and standard deviations
($\sigma_\Phi$). 
Both are significantly larger for the $\Phi_2$ angle compared
to $\Phi_1$. 
As in Fig.~\ref{fig06} for comparison we display two
theoretical angle distributions, Cauchy and uniform. 
The $\Phi_2$ distribution for smaller angles is shifted toward 
zero, compared to that in Fig.~\ref{fig06}; this effect may be 
caused by higher random fluctuations of the observed angle.

In Fig.~\ref{fig08} we display a scatterplot of two angles
$\Phi_1$ and $\Phi_2$.
A relatively high correlation coefficient ($\rho=0.44$)
implies that the focal mechanism forecast performs reasonably
well, and its uncertainty is reasonably well evaluated by the
angle $\Phi_1$. 
However, the distribution of either angles is not Gaussian,
hence the correlation coefficient and regression parameters
should be considered with a certain caution.
However, at least the distribution offers some quantitative
measure of angles mutual dependence. 
Therefore, we need to carry out additional testing with 
modified forecast parameters to determine appropriate measure 
of the forecast skill. 
This topic needs to be investigated in future studies.

Figs.~\ref{fig09} and \ref{fig10} show the result of applying
the bearing angle correction (Eqs.~\ref{eq14}--\ref{eq20})
to estimate the $\Phi_1$ angle.
The corrected angle is slightly larger for latitudes
approaching polar areas.
As we mentioned earlier, this angle depends on earthquake 
spatial distribution around a forecast point, as well as 
point's proximity to the pole.
Additional studies, perhaps involving simulated earthquake 
spatial distribution, need to be carried out to understand 
these features of the $\Phi_1$ angle distribution.
On the other hand, for the $\Phi_2$ angle the bearing
correction result is opposite (see Figs.~\ref{fig11} and
\ref{fig12}): after the correction the observed mechanism is
in a better agreement with the predicted focal mechanism. 

After solving the problem of bearing correction for polar
areas, we extend our focal mechanism forecast to
$[ 90^\circ S - 90^\circ N ]$ latitude range, i.e., the whole
Earth. Close to the poles we need to use the exact spherical
distance formula (for example, Turner 1914; Bullen 1979,
Eq.~5, p.~155) which requires about twice the computation
time.

Kagan \& Jackson (2012) performed such whole Earth forecast
based on the PDE (Preliminary determinations of
epicenters) catalog (PDE 2012).
No prediction of the focal
mechanism was done in this work, because the PDE catalog lacks
focal mechanism estimate for many earthquakes. 
However, the PDE catalog contains many smaller earthquakes, 
and its magnitude threshold is $m_t=5.0$, so the forecast 
has a better spatial resolution.

\section{Source complexity}
\label{comp}

There are several ways to measure earthquake source complexity.
The simplest method, which can be applied to a single
earthquake seismic moment tensor, is the CLVD $\Gamma$-index 
(Kagan \& Knopoff 1985a).
The $\Gamma$-index equals zero for a double-couple source,
and its value $+1$ or $-1$ corresponds to the pure CLVD
mechanism of the opposite sign. 

In Fig.~\ref{fig13} we display the index distribution for the
GCMT catalog.
The distribution is concentrated around the $\Gamma$-value
close to zero, i.e., most earthquakes have a double-couple
focal mechanism or a focal mechanism that is close to
double-couple.
Kagan (2002, his Fig.~6) obtained a similar estimate of the
$\Gamma$-index standard deviation ($\sigma_\Gamma$) dependence
on magnitude. 

The $\Gamma$-index standard deviation (0.39) shown in
Fig.~\ref{fig13} is significantly smaller than that of the
uniform distribution: 
\be
\sigma^u_\Gamma \ = \ 2/\sqrt 3 \ = \ 1.155
\, .
\label{eq22}
\ee
Kagan \& Knopoff (1985a) showed that for the sum of randomly
rotated focal mechanisms the distribution of the
$\Gamma$-index is uniform at the range [$-1,~1$]. 
However, for tectonic events non-$DC$ mechanisms like
the CLVD are likely due to various systematic and random
errors in determining the mechanism (Frohlich \& Davis 1999;
Kagan 2003).
These results suggest that routinely determined CLVD values
would not reliably show the deviation of earthquake focal
mechanisms from a standard $DC$ model.

Kagan (2000, Figs.~4a and 5a) simulated the effect of reported
moment inversion errors in the GCMT catalog on possible values
of the $\Gamma$-index and found that these inversion errors
may cause significant standard errors, up to 0.2, in the
$\Gamma$-index. 
Kagan (2000, 2003) suggests that the internal uncertainties in
the GCMT data may be only a part of the total random and
systematic errors.
Therefore, it is quite feasible that $\sigma_\Gamma = 0.39$
shown in Fig.~\ref{fig13} is due to these systematic and 
random errors. 

Several techniques have been proposed for measuring the
complexity of focal mechanism distribution in a region. 
Kagan \& Knopoff (1985b) suggested measuring irregularity of
the earthquake focal mechanism distribution by calculating
three scalar invariants of a moment tensor set.
The simplest invariant is
\be
J_3 \ = \ < {\bf m}_{ij} \, {\bf n}_{ij} > \, ,
\label{eq25}
\ee
where $<>$ is the averaging symbol; ${\bf m}_{ij}$ and ${\bf
n}_{ij}$ are earthquake moment tensors.
The standard index summation is assumed.
For normalized tensors $ -2 \le J_3 \le 2$.
The former equality characterizes the oppositely rotated
tensors and the latter equality the equally oriented tensors
(see also Alberti 2010). 

Apperson (1991, Eq.~4) recommends characterizing complexity for
a group of earthquakes by a ``seismic consistency" index
($C_s$) which is the ratio of the summed seismic moment tensor
for $n$ earthquakes to a sum of their scalar moments:
\be
C_s \ = \ { | {\sum\limits_1^n {\bf m} | } \over
{\sum\limits_1^n M }} \, , 
\label{eq26}
\ee
where $M$ is the scalar moment, $\bf m$ is the tensor, and 
$| \, |$ means that the scalar moment of the tensor sum is
taken. 
For earthquakes with identically oriented focal mechanisms
$C_s = 1$; for randomly disoriented sources $C_s = 0$.
Bailey {\it et al.} (2010) used $C_s$ to investigate the
complexity of the focal mechanism distribution in
California.

In our forecast of focal mechanisms we calculate the seismic
moment tensor for all events within a 1000~km circular area
around each forecast point by using Eq.~\ref{eq3}. 
For this averaged moment tensor, we calculate the
$\Gamma$-index as well as the average rotation angle $\Phi_1$
between the forecasted double-couple and all other earthquakes
in the 1000~km circular area. 
These two variables indicate the complexity of the focal
mechanism distribution around the forecast point. 
The advantage of these two complexity measures is that the
angle $\Phi_1$ has a clear geometrical meaning and can be 
roughly evaluated by inspection of focal mechanism maps.
The $\Gamma$-index general and statistical properties are
known (see above). 

Fig.~\ref{fig14} displays a two-dimensional distribution of
the $\Gamma$-index vs the forecasted angle $\Phi_1$.
For large $\Gamma$-values the angle is also large, whereas
for small $\Gamma$ the angle can be practically of any
value ($ 0^\circ \le \Phi_1 \le 120^\circ$).
In Fig.~\ref{fig16} a two-dimensional distribution is
shown for predicted 2008-2012 earthquakes.
The distributions in Figs.~\ref{fig14} and \ref{fig16} can be
expected to be similar. 
They are obtained by the same computational procedure.
A similar picture for the ``seismic consistency" index
(\ref{eq26}) is shown by Bailey {\it et al.} (2010, their
Fig.~3). 
However, Fig.~\ref{fig17} displays a different behaviour: the
$\Phi_2$ angle points are scattered over the diagram: the
observed focal mechanisms are less correlated with the CLVD
component of the predicted moment tensor.

In Figs.~\ref{fig19} and \ref{fig20} we show marginal
distributions of the angle $\Phi_1$ and $\Gamma$-index
for all forecasted cells.
About 42\% of the cells have these variables equal to zero.
These cell centers do not have any earthquake centroid within
1000~km distance.
To avoid future `surprises', we assume 1\% of all earthquakes
to occur uniformly over the Globe (Jackson \& Kagan 1999;
Kagan \& Jackson 2000). 
These places can be identified in Fig.~\ref{fig04} by 
the greenish-gray color. 
In Fig.~\ref{fig14} these zero values of $\Phi_1$ and $\Gamma$
are all plotted at the point [0.0,~0.0] (see Eq.~\ref{eq21})
and thus are not visible. 

The CLVD source can be decomposed in various ways (Wallace
1985; Julian {\it et al.}\ 1998).
If we arrange the moment tensor eigenvalues in their absolute
values as $|\lambda_1| \ge |\lambda_2| \ge |\lambda_3|$, then
in our forecast we use $|\lambda_1|, |\lambda_2|$, depending
on their sign, as the $T$- or $P$-axes of a double-couple, and
we assign $|\lambda_1|$ eigenvalue with an opposite sign to
these axes. 
This corresponds to the selection of the major double-couple
as a representative of the predicted focal mechanism (Wallace
1985, his Eq.~3; Julian {\it et al.}\ 1998, their Fig.~3b).
The remaining part of the moment tensor is called a minor
double-couple.
Though we do not list the minor double-couple in our forecast
table, it can be easily done.
On the other hand, we could explore other decompositions
(Wallace 1985; Julian {\it et al.}\ 1998) of the predicted
seismic moment tensor, and, in particular, we can study what
predictive skill they may have.

\section{Discussion }
\label{disc}

Although in our previous forecasts (Kagan \& Jackson 1994,
2000; Jackson \& Kagan 1999) we optimized the smoothing kernel
to obtain a better prediction of the future seismicity rate,
focal mechanism forecasts were not specifically optimized.
In Fig.~\ref{fig08} we showed how this optimization can be
accomplished.
As Eq.~\ref{eq3} indicates, several parameters can be
involved in the optimization, making it a time-consuming task.

Until now in our forecasts we have predicted only the
long-term focal mechanisms. 
However, a similar technique can be applied to the short-term
forecast prediction.
Kagan (2000) investigated the temporal correlations of
earthquake focal mechanisms and showed that at short time
intervals future focal mechanisms closely follow the
mechanisms of recent earthquakes.
These results can be applied to evaluate short-term forecasts
of focal mechanisms.

Moreover, our forecasts of short-term seismicity rates are
largely based on the results of the likelihood analysis of
earthquake catalogs (Jackson \& Kagan 1999) and Kagan \&
Jackson 2000; 2011).
In such analysis only space-time patterns of earthquake
occurrence have been investigated; focal mechanisms were not
included.
Both long- and short-term forecasts can certainly be improved
by a method which would incorporate focal mechanism 
similarity in the likelihood calculation.

In our forecasts of focal mechanisms we need to resolve the
fault-plane ambiguity, i.e., to decide which of two focal
planes in the GCMT moment tensor solution is a fault-plane. 
This information is necessary to extend our rate forecast
along the fault-plane (Eq.~\ref{eq2}).
This equation also governs the selection of earthquake focal
mechanisms to infer predicted moment tensor (Eq.~\ref{eq3}).
In the GCMT catalog such a decision is based on a statistical
guess (Kagan \& Jackson, 1994, their Fig.~3) which is correct
only in about 75\% cases in subduction zones. 
In other tectonic regions it is likely that the fault-plane
guess selection is correct only in about 50\% cases.

In many cases, especially in continental areas, additional
geological and seismic information exists which might help
resolve the fault-plane ambiguity (Kagan {\it et al.}\ 2006;
Wang {\it et al.}\ 2009).
Moreover, aftershock pattern (Kagan 2002) and surface
deformation measurements can also supply necessary
information.
Such a program would require a significant but feasible work.

Bird {\it et al.}\ (2010a) used a global strain rate model 
(GSRM) to construct a high-resolution forecast based on
moment-rates inferred from geodetic strain rate data. 
Focal mechanisms can also be estimated from the observed
geodetic strain rate tensor. 
In most cases the geodetic data constrain only the horizontal
components, so that some additional data or assumptions
regarding the dip angle are needed to estimate the full strain
rate tensor. 
As for seismic moment tensors, the major and minor
double-couples each have two nodal planes, and additional data
or assumptions are needed to determine which corresponds to
the fault plane. 
In developing work, Bird {\it et al.}\ (2010b) have found that
hybrid forecasts combining smoothed seismicity and strain rate
forecasts performed better than either one by itself. 
We anticipate that the same will hold for focal mechanism
forecasts. 

Quasi-static Coulomb stress provides another possible tool for
earthquake forecasting (Stein, 1999; King {\it et al.}, 1994;
Toda \& Stein 2003).
The primary hypothesis is that changes in Coulomb stress
caused by a ``source" earthquake, resolved onto the rupture
plane of a future ``receiver" earthquake, brings that fault
closer to failure. 
In retrospective testing, one can know the rupture plane, but
for prospective forecasting the eventual rupture plane, onto
which the tensor stress should be resolved, is unknown. 
Focal mechanism forecasts, as described above, can provide
most probable options in the form of the two nodal planes of
the double-couple moment tensors. 
Additional data may in some cases indicate which is more
likely to be the fault rupture plane. 

\subsection* {Acknowledgments
}
\label{Ackn}

\begin{figure}
\begin{center}
\includegraphics[width=0.85\textwidth,angle=-90]{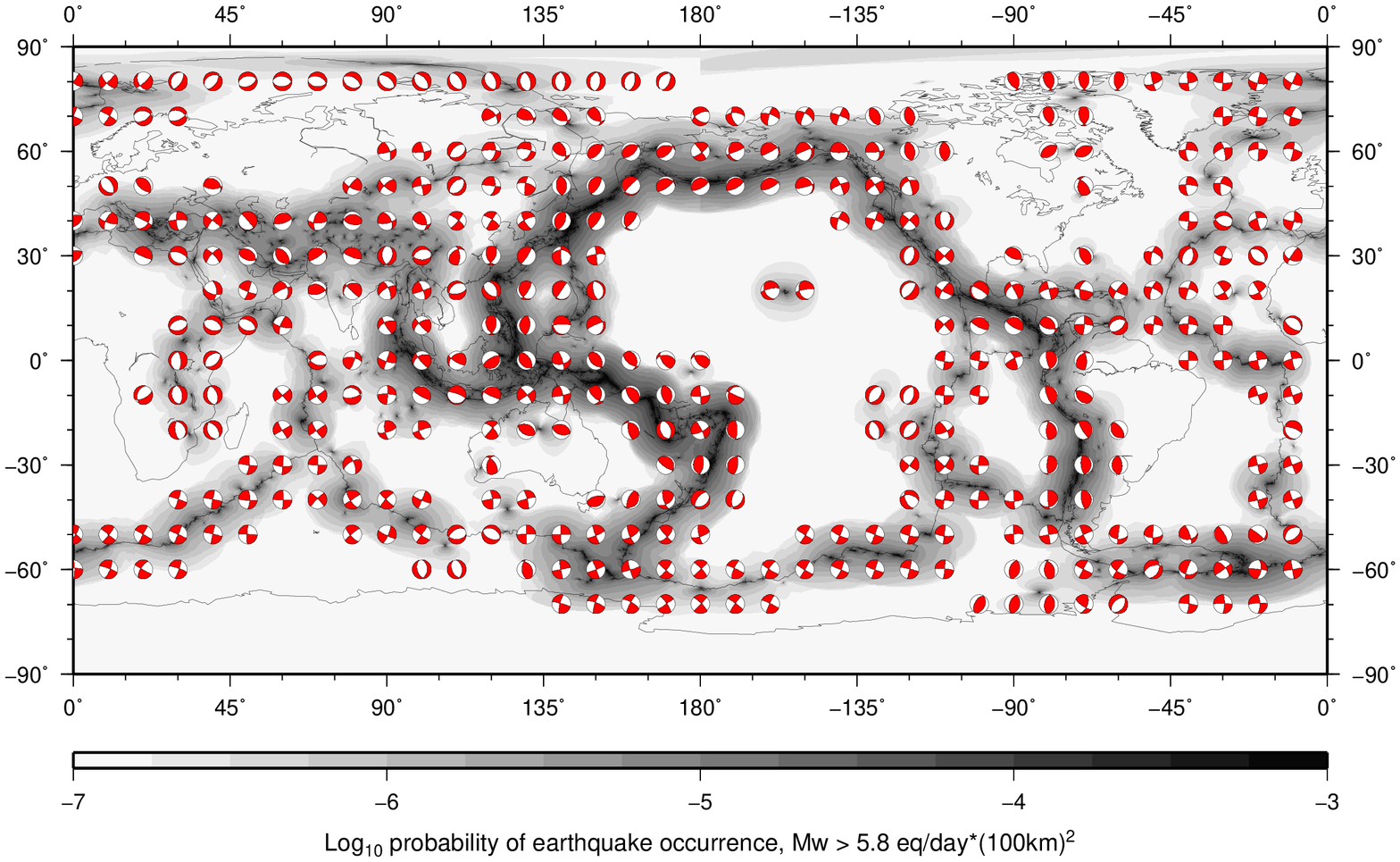}
\caption{Global earthquake long-term focal mechanism forecast based on
}
\label{fig04a}
\end{center}
\vskip -.5cm
smoothed seismicity, latitude range $[ 90^\circ S-90^\circ N
]$.
Focal mechanisms are shown on $5^\circ \times 5^\circ $ grid.
\end{figure}

\begin{figure}
\begin{center}
\includegraphics[width=0.65\textwidth]{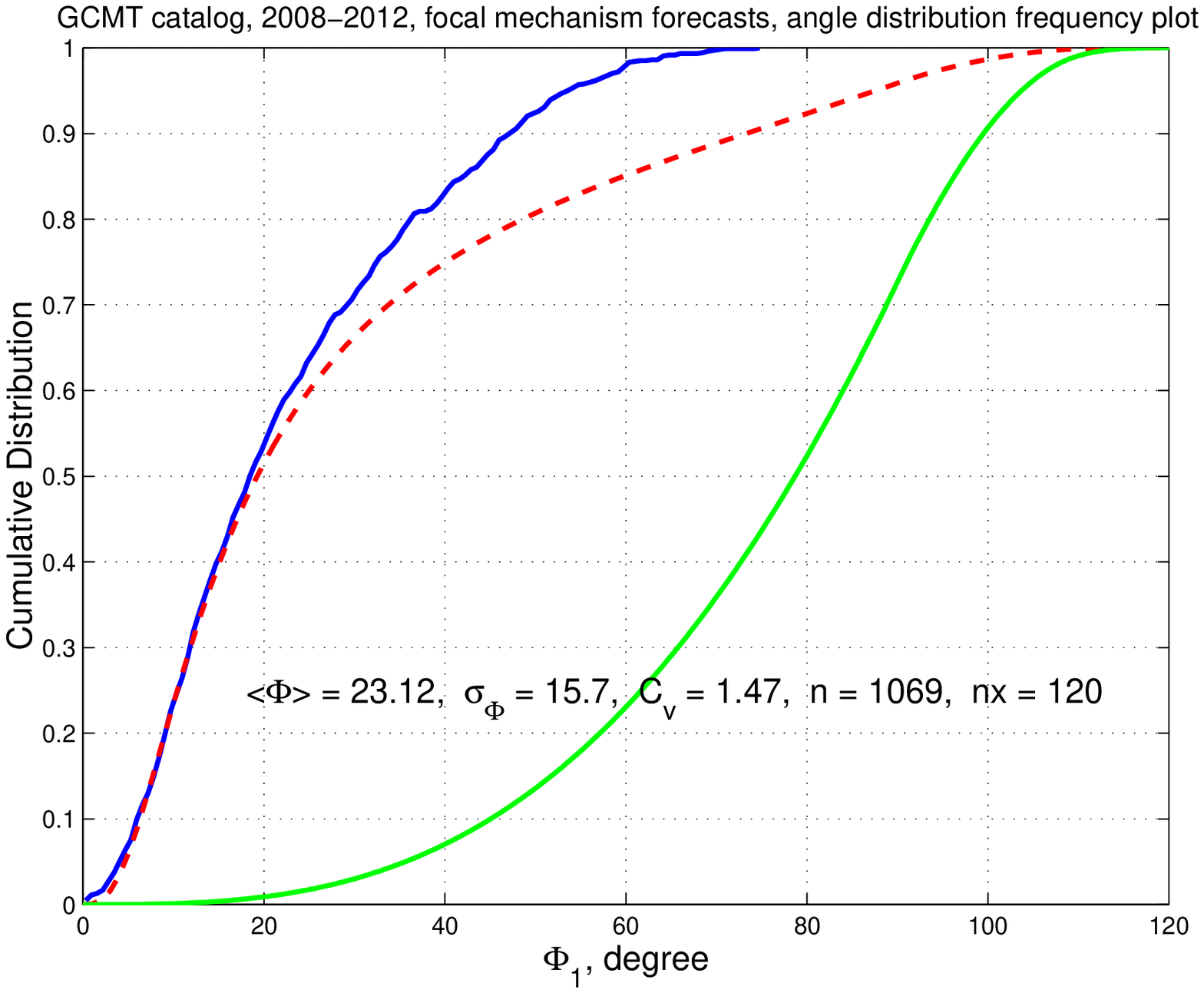}
\caption{GCMT catalog, 2008--2012, earthquake number 
$n=1069$. 
} 
\label{fig06}
\end{center}
\vskip -.5cm
Blue curve is cumulative distribution of predicted rotation
angle $\Phi_1$ at earthquake centroids. 
The red dashed line is for the Cauchy rotation with $\kappa =
0.075$. 
Right green solid line is for the random rotation.
\end{figure}

\begin{figure}
\begin{center}
\includegraphics[width=0.65\textwidth]{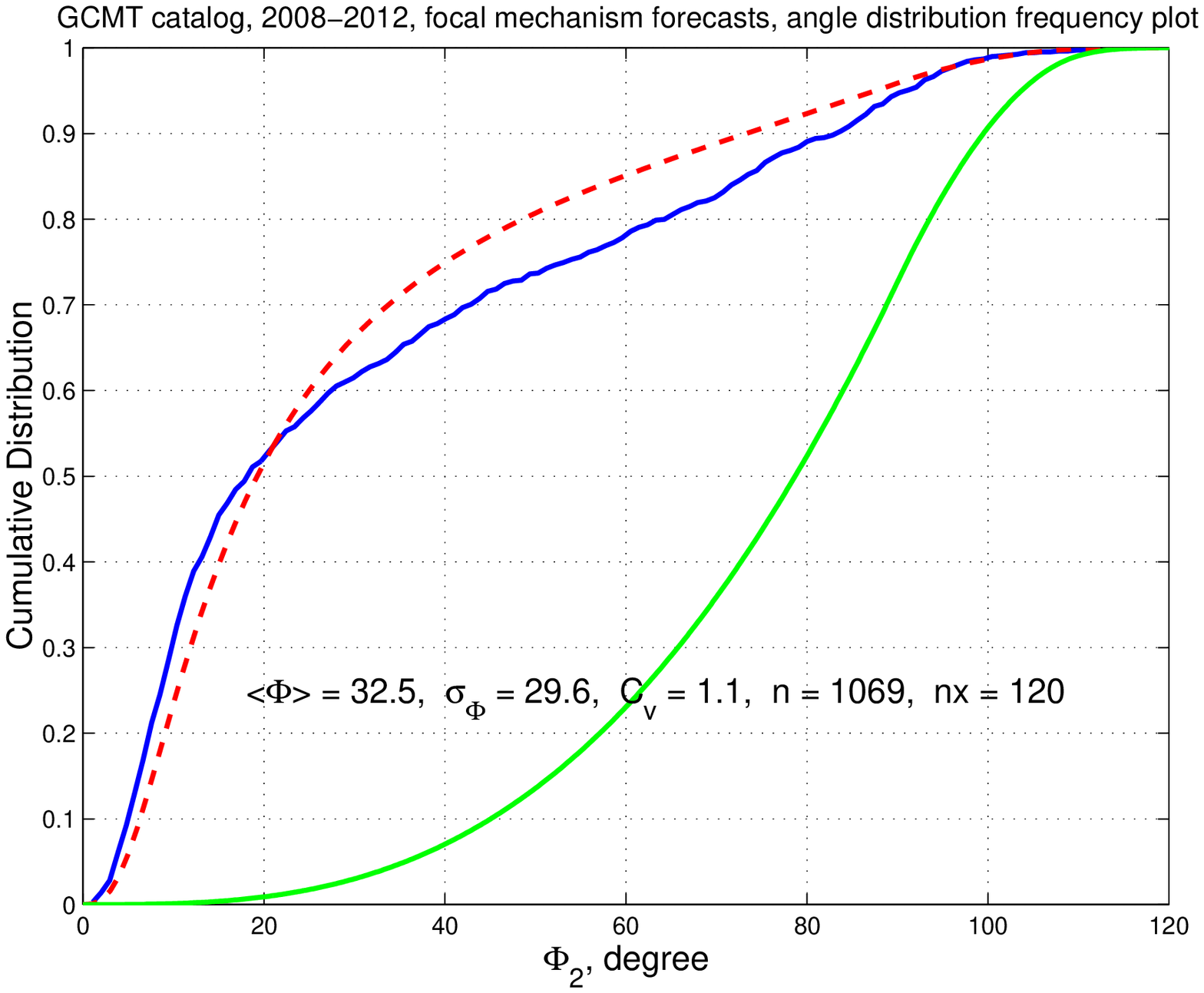}
\caption{GCMT catalog, 2008--2012, earthquake number 
$n=1069$. 
} 
\label{fig07}
\end{center}
\vskip -.5cm
Blue curve is cumulative distribution of observed rotation
angle $\Phi_2$. 
The red dashed line is for the Cauchy rotation with $\kappa =
0.075$. 
Right green solid line is for the random rotation.
\end{figure}

\begin{figure}
\begin{center}
\includegraphics[width=0.65\textwidth]{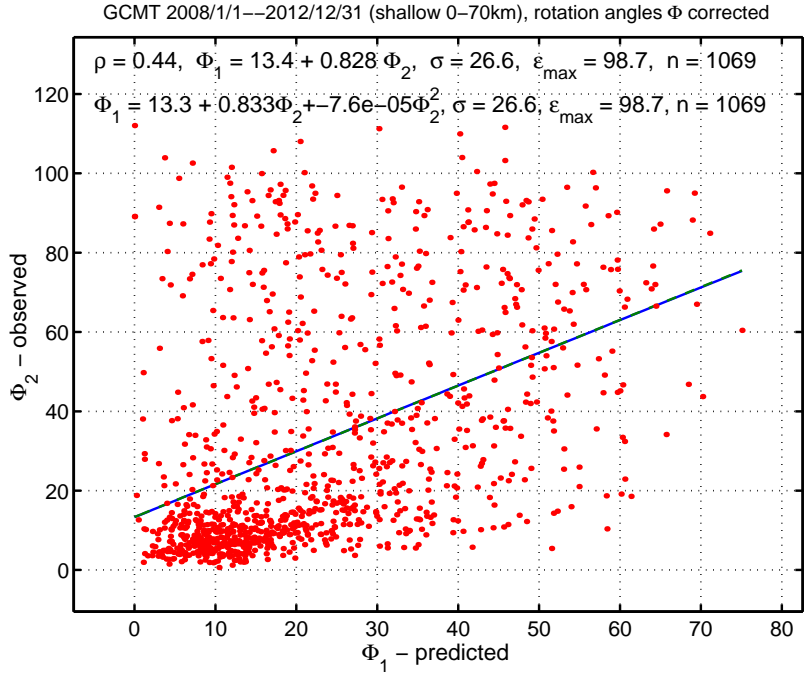}
\caption{Distribution of rotation angles }
\label{fig08}
\end{center}
\vskip -.5cm
in the Global Centroid Moment Tensor (GCMT) catalog,
1977--2012, earthquake number $n=1069$. 
We calculate two regression lines approximating the
interdependence of the predicted $\Phi_1$ and observed
$\Phi_2$ angles, the linear and quadratic curves. 
The curves overlap, testifying that the linear regression fits.
The coefficient of correlation between the angles is $0.44$,
indicating that the $\Phi_1$ estimate forecasts the uncertainty
for future mechanisms reasonably well.
\end{figure}

\begin{figure}
\begin{center}
\includegraphics[width=0.65\textwidth]{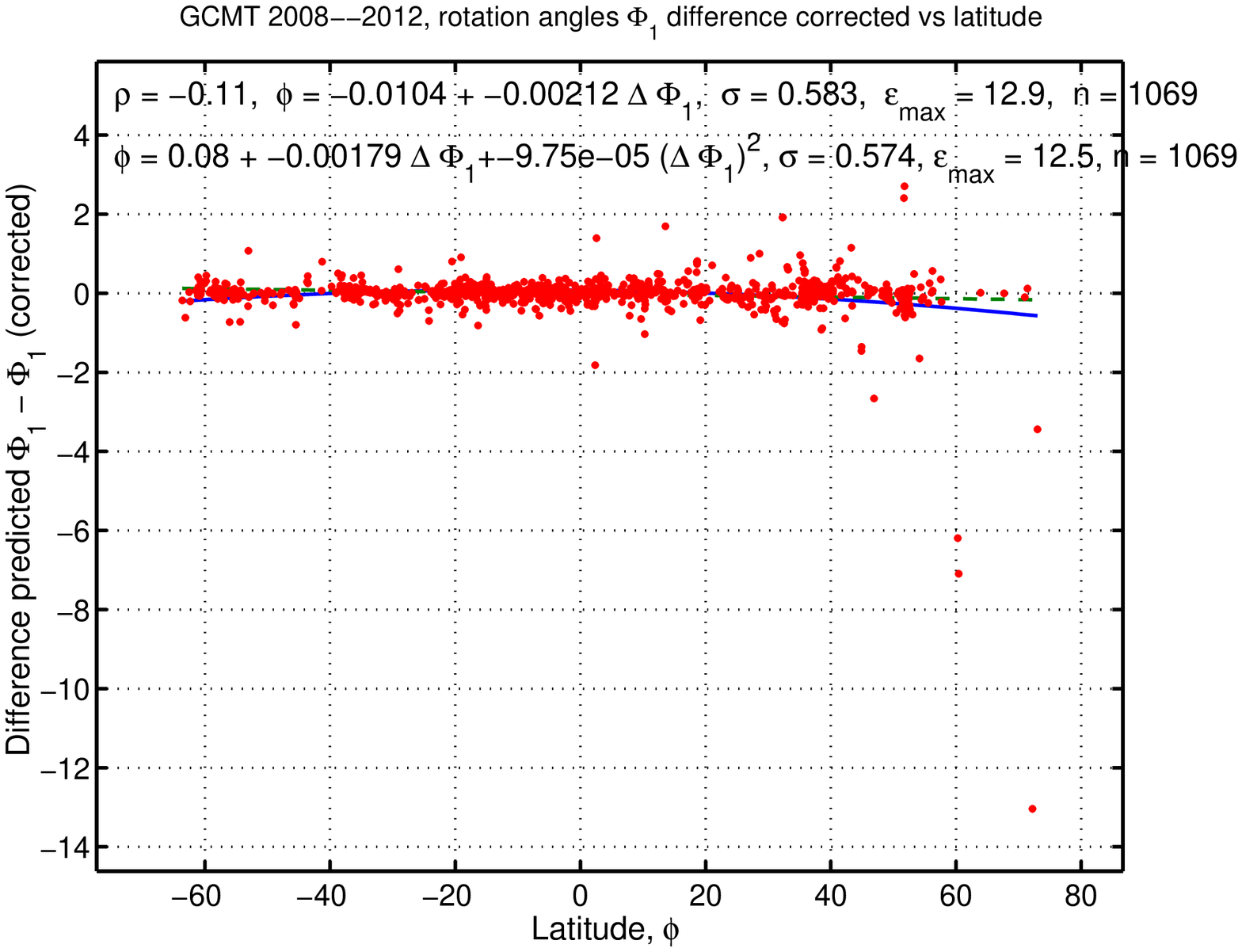}
\caption{Distribution of difference of predicted rotation
angles $\Phi_1$ 
} 
\label{fig09}
\end{center}
\vskip -.5cm
in the original program (Kagan \& Jackson, 2011) and
with the angle corrected according to
Eqs.~\ref{eq14}--\ref{eq20}. 
\end{figure}

\begin{figure}
\begin{center}
\includegraphics[width=0.65\textwidth]{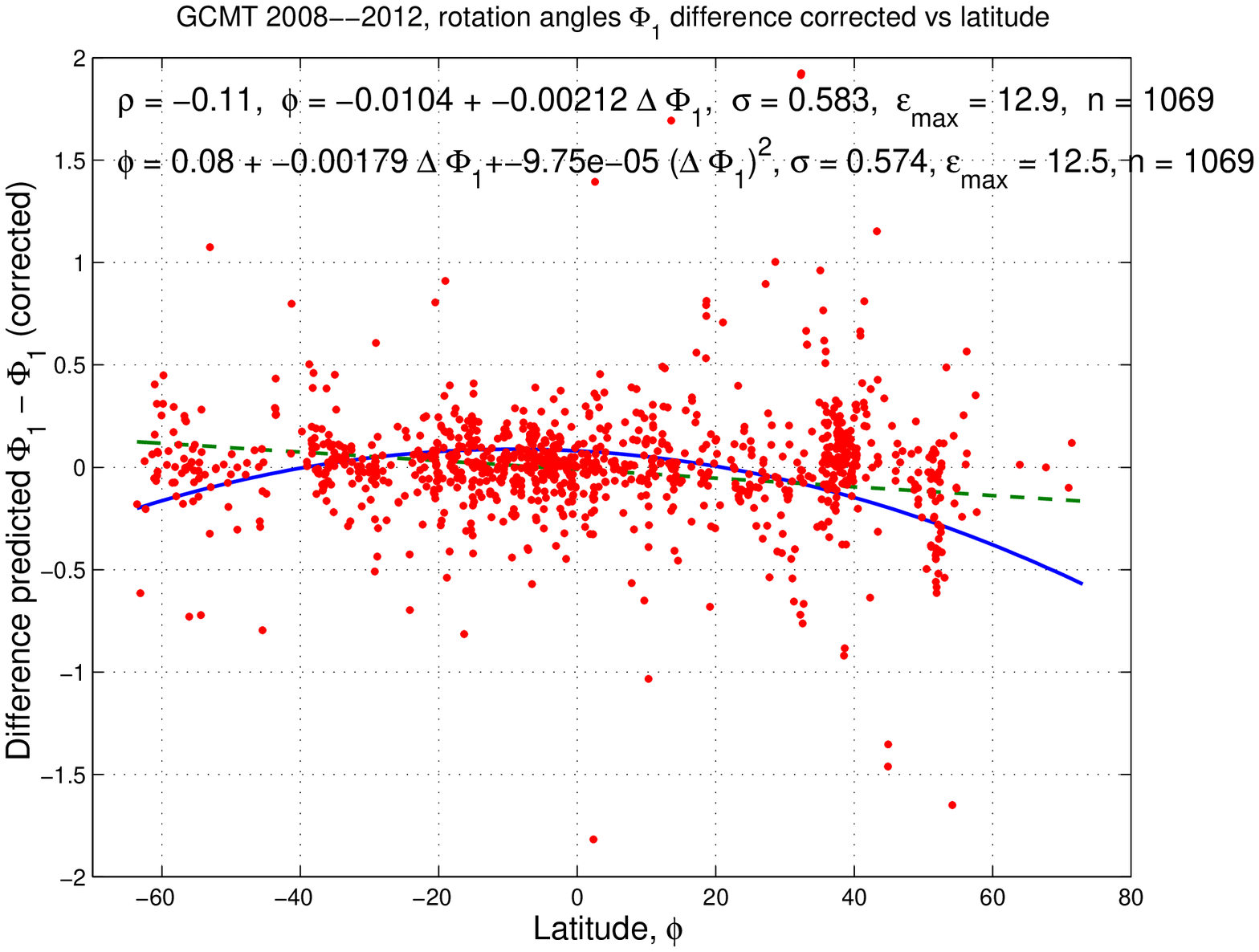}
\caption{Same as Fig.~\ref{fig09} with the vertical scale
expanded. 
}
\label{fig10}
\end{center}
\vskip -.5cm
\end{figure}

\begin{figure}
\begin{center}
\includegraphics[width=0.65\textwidth]{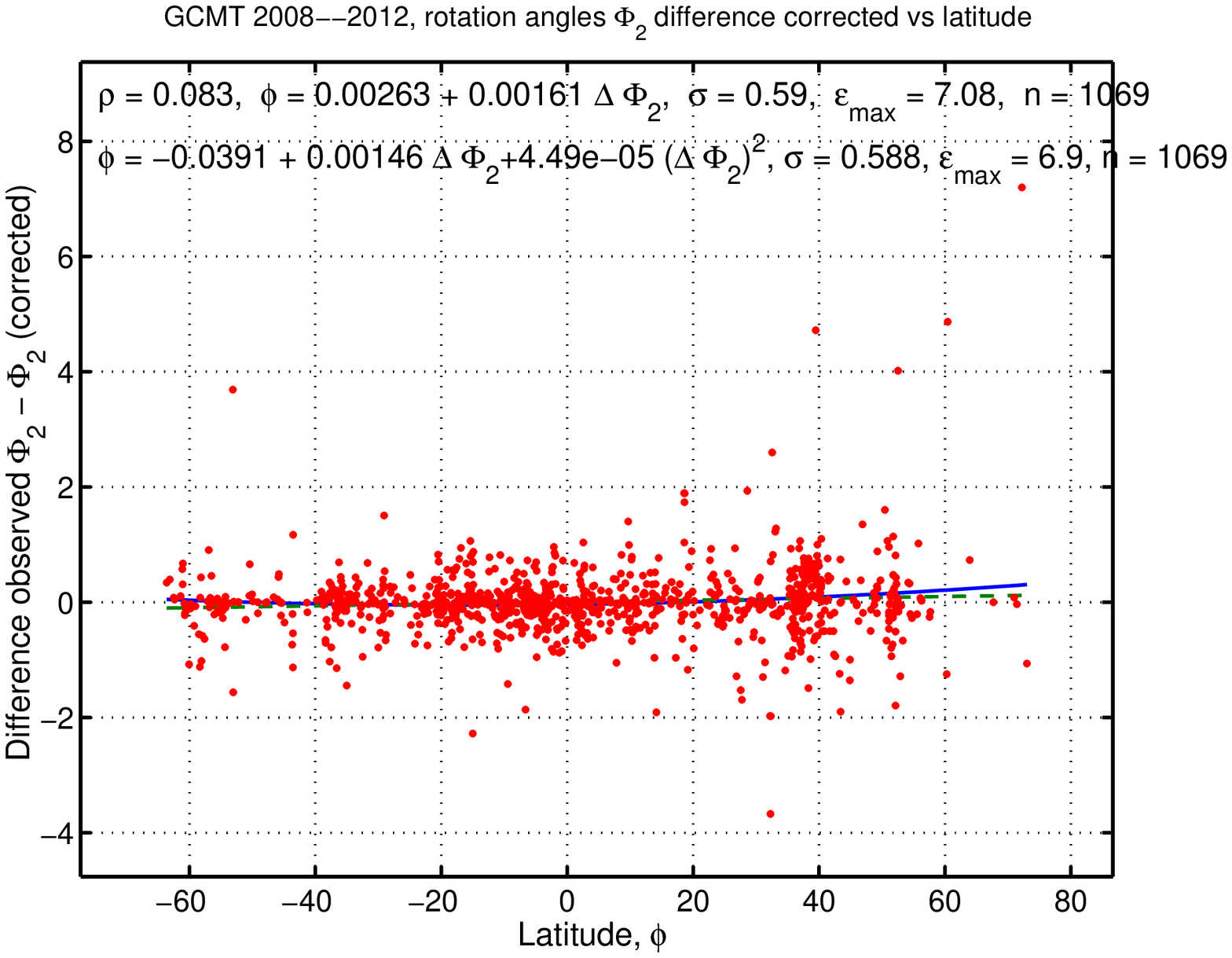}
\caption{Distribution of difference of observed rotation
angles $\Phi_2$ 
} 
\label{fig11}
\end{center}
\vskip -.5cm
in the original program (Kagan \& Jackson, 2011) and
the angle corrected according to Eqs.~\ref{eq14}--\ref{eq20}.
\end{figure}

\begin{figure}
\begin{center}
\includegraphics[width=0.65\textwidth]{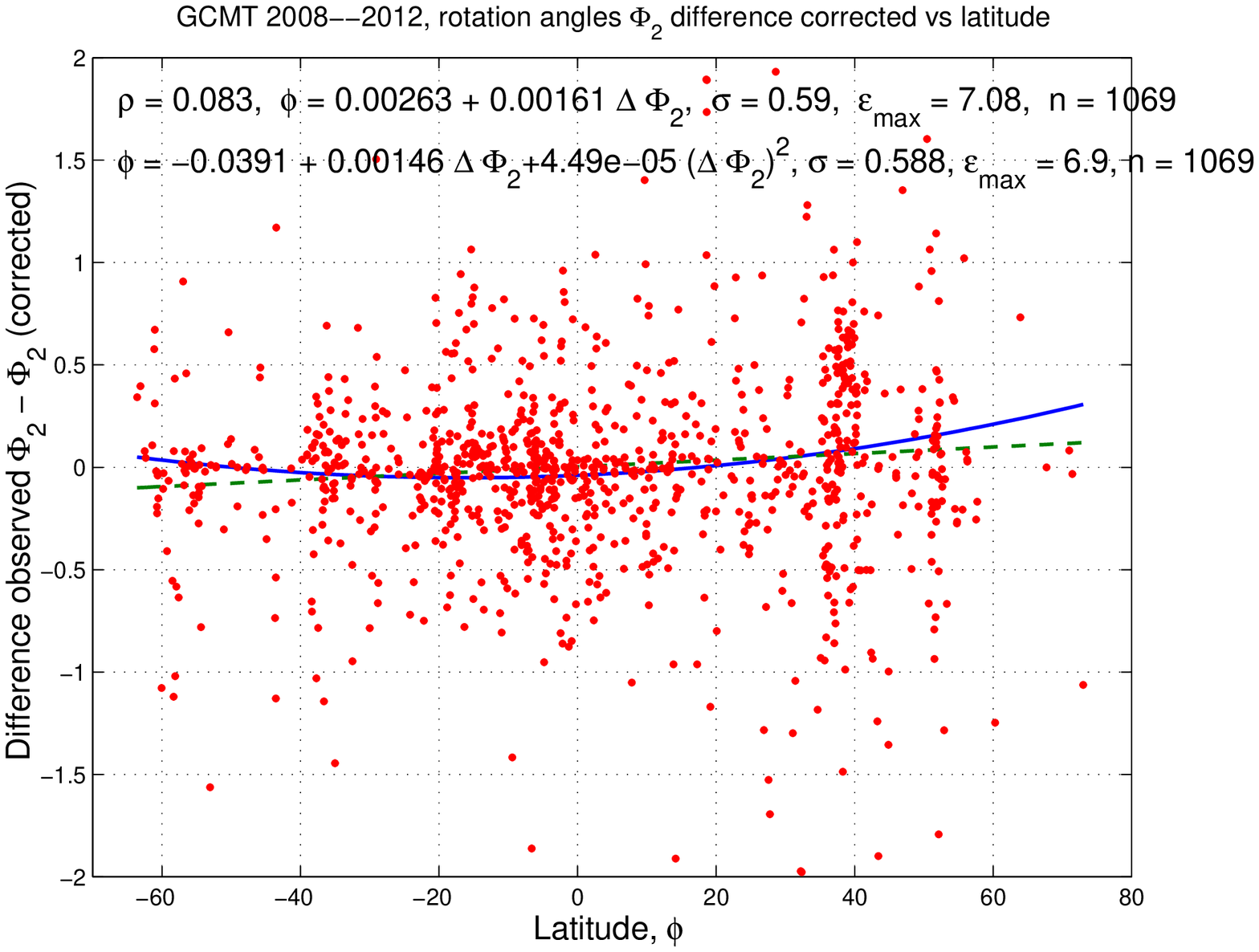}
\caption{Same as Fig.~\ref{fig11} with the vertical scale
expanded. 
}
\label{fig12}
\end{center}
\vskip -.5cm
\end{figure}

\begin{figure}
\begin{center}
\includegraphics[width=0.65\textwidth]{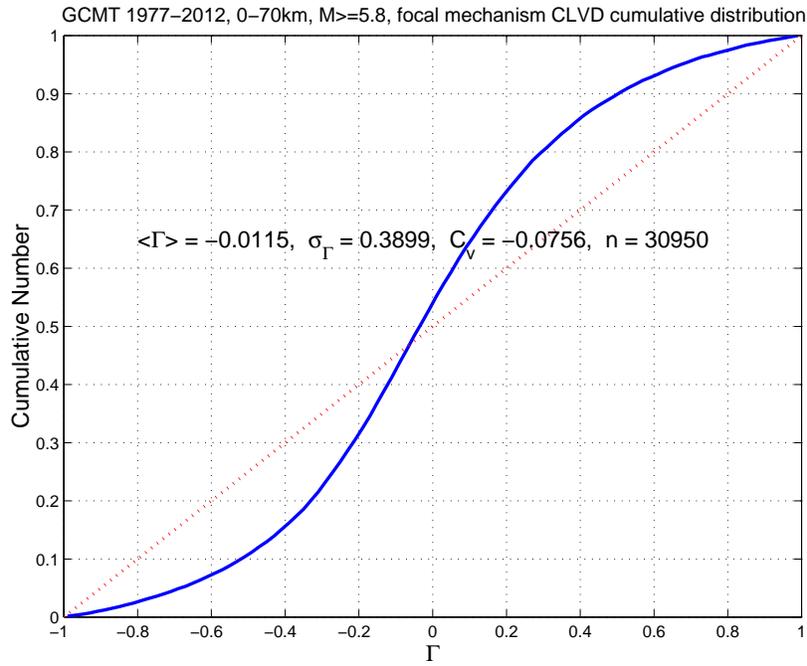}
\caption{Blue solid line -- cumulative distribution of
$\Gamma$ index for seismic moment tensor of shallow
earthquakes in the GCMT catalog, 1977-2012, $m_w \ge 5.8$. 
Red dotted line is the uniform distribution of $\Gamma$.
}
\label{fig13}
\end{center}
\vskip -.5cm
\end{figure}

\begin{figure}
\begin{center}
\includegraphics[width=0.65\textwidth]{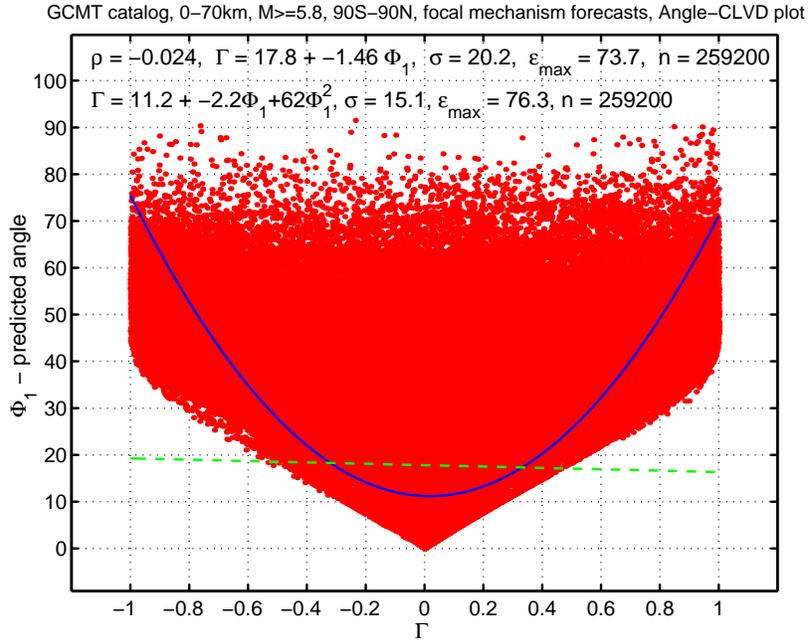}
\caption{Scatterplot of $\Gamma$ index vs forecasted angle
$\Phi_1$ for all cells in 90$^\circ$S -- 90$^\circ$N
forecast.
}
\label{fig14}
\end{center}
\vskip -.5cm
\end{figure}

\begin{figure}
\begin{center}
\includegraphics[width=0.65\textwidth]{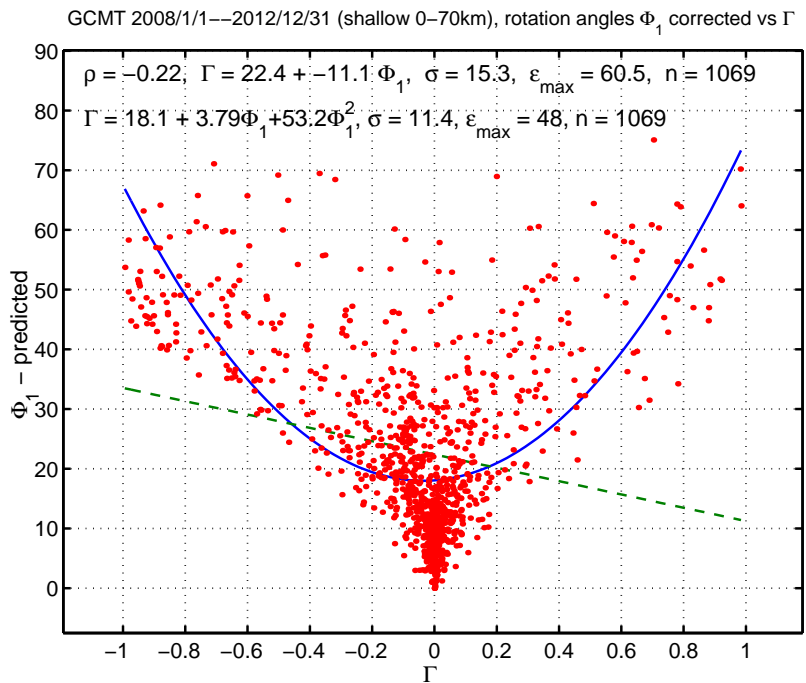}
\caption{Scatterplot of $\Gamma$ index vs forecasted angle
$\Phi_1$ for 2008-2012 earthquakes.
}
\label{fig16}
\end{center}
\vskip -.5cm
\end{figure}

\begin{figure}
\begin{center}
\includegraphics[width=0.65\textwidth]{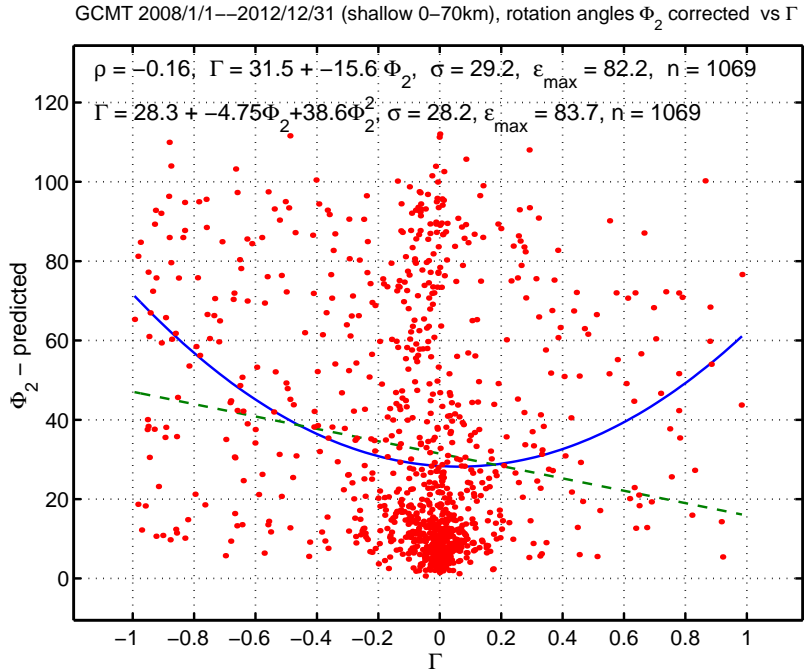}
\caption{Scatterplot of $\Gamma$ index vs observed angle
$\Phi_2$ for 2008-2012 earthquakes.
}
\label{fig17}
\end{center}
\vskip -.5cm
\end{figure}

\begin{figure}
\begin{center}
\includegraphics[width=0.65\textwidth]{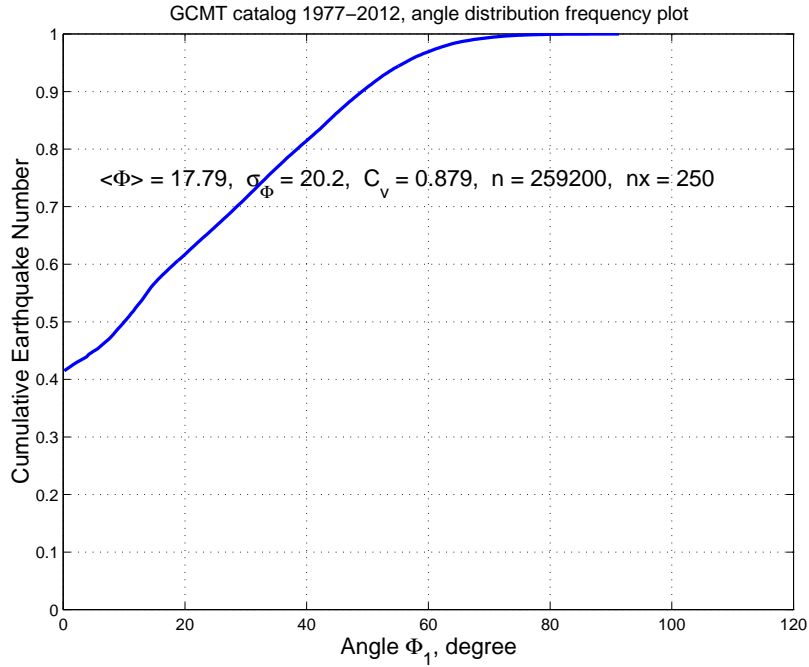}
\caption{Cumulative distribution of forecasted angle
$\Phi_1$ for all 
cells in 90$^\circ$S -- 90$^\circ$N
forecast. 
}
\label{fig19}
\end{center}
\vskip -.5cm
\end{figure}

\begin{figure}
\begin{center}
\includegraphics[width=0.65\textwidth]{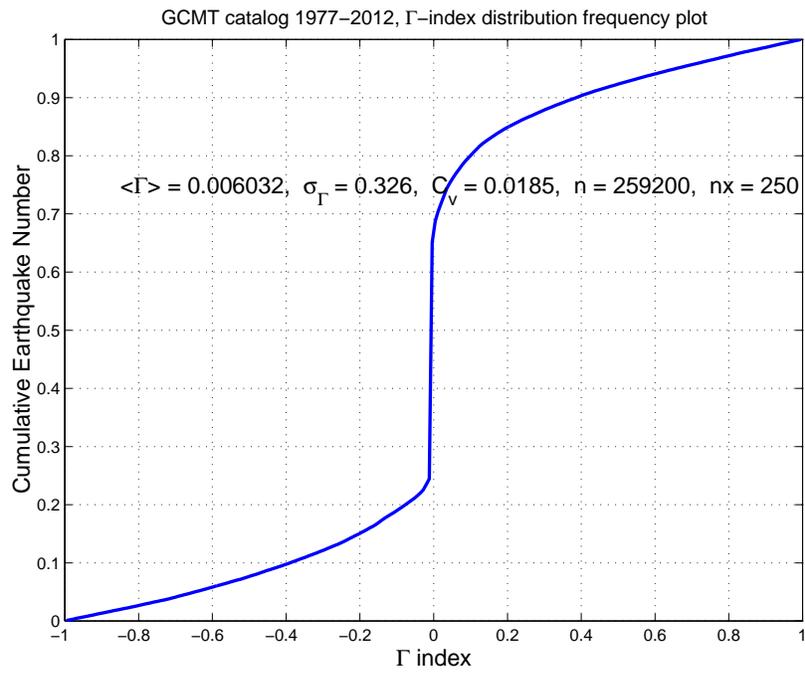}
\caption{Cumulative distribution of $\Gamma$-index
for all 
cells in 90$^\circ$S -- 90$^\circ$N forecast. 
}
\label{fig20}
\end{center}
\vskip -.5cm
\end{figure}

We are grateful to Peter Bird of UCLA for useful discussion
and suggestions as well as to Kaiqing Yuan and Igor Stubailo
for help in GMT plotting. 
Fig.~\ref{fig04a} was prepared using the Generic Mapping Tools
(Wessel \& Smith 1998). 
The authors appreciate support from the National Science
Foundation through grants EAR-0711515, EAR-0944218, and
EAR-1045876.

\pagebreak

\centerline { {\sc References} }
\vskip 0.1in
\parskip 1pt
\parindent=1mm
\def\reference{\hangindent=22pt\hangafter=1}

\reference
Aki, K.~\& P. G. Richards, 2002.
{\sl Quantitative Seismology},
2nd ed., Sausalito, Calif., University Science Books, 700~pp.

\reference
Alberti, M., 2010.
Analysis of kinematic correlations in faults and focal
mechanisms with GIS and Fortran programs,
{\sl Computers and Geosciences}, {\bf 36}(2), 186-194.

\reference
Apperson, K. D., 1991.
Stress Fields of the Overriding Plate at Convergent Margins
and Beneath Active Volcanic Arcs,
{\sl Science}, {\bf 254}(5032), 670-678.

\reference
Bailey, I. W., Ben-Zion, Y., Becker, T. W. \&
Holschneider, M., 2010.
Quantifying focal mechanism heterogeneity for fault
zones in central and southern California,
{\sl Geophys.\ J. Int.}, {\bf 183}(1), 433.450.

\reference
Bird, P., C. Kreemer \& W. E. Holt, 2010a.
A long-term forecast of shallow seismicity based on the
Global Strain Rate Map,
{\sl Seismol.\ Res.\ Lett.}, {\bf 81}(2), 184-194
(plus electronic supplement).

\reference
Bird, P., Y. Y. Kagan \& D. D. Jackson, 2010b.
Time-dependent global seismicity forecasts with a tectonic
component: Retrospective tests,
AGU Fall Meet.\ Abstract S44B-03.

\reference
Bullen, K. E., 1979.
{\sl An Introduction to the Theory of Seismology},
3rd ed., Cambridge, Cambridge University Press, 381~pp.

\reference
Eberhard, D. A. J., J. D. Zechar \& S. Wiemer, 2012.
A prospective earthquake forecast experiment in the western 
Pacific,
{\sl Geophys.\ J. Int.}, {\bf 190}(3), 1579-1592.

\reference
Ekstr\"om, G., M. Nettles \& A.M. Dziewonski, 2012.
The global CMT project 2004-2010: Centroid-moment tensors for
13,017 earthquakes,
{\sl Phys.\ Earth Planet.\ Inter.}, {\bf 200-201}, 1-9.

\reference
Frohlich, C. \& S.\ D.\ Davis, 1999.
How well constrained are well-constrained T, B, and P axes in
moment tensor catalogs?,
{\sl J.\ Geophys.\ Res.}, {\bf 104}, 4901-4910.

\reference
Gasperini, P. \& G. Vannucci 2003.
FPSPACK: a package of FORTRAN subroutines to manage earthquake 
focal mechanism data,
{\sl Computers \& Geosciences}, {\bf 29}(7), 893-901, DOI:
10.1016/S0098-3004(03)00096-7.

\reference
Hanks, T. C., 1992.
Small earthquakes, tectonic forces,
{\sl Science}, {\sl 256}, 1430-1432.

\reference
Jackson, D.~D. \& Y.~Y.~Kagan, 1999.
Testable earthquake forecasts for 1999,
{\sl Seism.\ Res.\ Lett.}, {\bf 70}(4), 393-403.

\reference
Julian, B. R., A. D. Miller \& G. R. Foulger, 1998.
Non-double-couple earthquakes, 1. Theory,
{\sl Rev.\ Geophys.}, {\bf 36}(4), 525-549.

\reference
Kagan, Y.~Y., 1991.
3-D rotation of double-couple earthquake sources,
{\sl Geophys.\ J. Int.}, {\bf 106}(3), 709-716.

\reference
Kagan, Y. Y., 2000.
Temporal correlations of earthquake focal mechanisms,
{\sl Geophys.\ J. Int.}, {\bf 143}(3), 881-897.

\reference
Kagan, Y. Y., 2002.
Aftershock zone scaling,
{\sl Bull.\ Seismol.\ Soc.\ Amer.}, {\bf 92}(2), 641-655,

\reference
Kagan, Y. Y., 2003.
Accuracy of modern global earthquake catalogs,
{\sl Phys.\ Earth Planet.\ Inter.}, {\bf 135}(2-3),
173-209.

%
%
\reference
Kagan, Y.~Y. \& D.~D.~Jackson, 1994.
Long-term probabilistic forecasting of earthquakes,
{\sl J. Geophys.\ Res.}, {\bf 99}, 13,685-13,700.

\reference
Kagan, Y. Y. \& D. D. Jackson, 2000.
Probabilistic forecasting of earthquakes,
{\sl Geophys.\ J. Int.}, {\bf 143}, 438-453.

\reference
Kagan, Y. Y. \& Jackson, D. D., 2011.
Global earthquake forecasts,
{\sl Geophys.\ J. Int.}, {\bf 184}(2), 759-776.

\reference
Kagan, Y. Y. \& Jackson, D. D., 2012.
Whole Earth high-resolution earthquake forecasts,
{\sl Geophys.\ J. Int.}, {\bf 190}(1), 677-686.

\reference
Kagan, Y. Y., D. D. Jackson \& Y. F. Rong, 2006.
A new catalog of southern California earthquakes, 1800-2005,
{\sl Seism.\ Res.\ Lett.}, {\bf 77}(1), 30-38.

\reference
Kagan, Y.~Y. \& L. Knopoff, 1985a.
The first-order statistical moment of the seismic moment
tensor,
{\sl Geophys.\ J.\ Roy.\ astr.\ Soc.}, {\bf 81}(2), 429-444.

\reference
Kagan, Y.~Y. \& L. Knopoff, 1985b.
The two-point correlation function of the seismic moment
tensor,
{\sl Geophys.\ J. Roy.\ astr.\ Soc.}, {\bf 83}(3), 637-656.

\reference
Kanamori, H., 1977.
The energy release in great earthquakes,
{\sl J. Geophys.\ Res.}, {\bf 82}, 2981-2987.

\reference
King, G. C. P., Stein, R. S. \& J. Lin, 1994.
Static stress changes and the triggering of earthquakes,
{\sl Bull.\ Seismol.\ Soc.\ Amer.}, {\bf 84}, 935-953.

\reference
Kostrov, B.~V., 1974.
Seismic moment and energy of earthquakes, and seismic flow of
rock,
{\sl Izv.\ Acad.\ Sci.\ USSR, Phys.\ Solid Earth}, January,
13-21.

\reference
{\sl Preliminary determinations of epicenters (PDE)}, 2012.
U.S.\ Geological Survey, U.S.\ Dep.\ of Inter., Natl.\
Earthquake Inf.\ Cent.,
\hfil\break
http://neic.usgs.gov/neis/epic/epic.html and
\hfil\break
http://neic.usgs.gov/neis/epic/code\_catalog.html
(last accessed December 2012).

\reference
Richardus, P. \& R. K. Adler, 1972.
{\sl Map projections for geodesists, cartographers and
geographers},
Amsterdam, North-Holland Pub. Co., pp.~174.

\reference
Snoke, J. A.\ 2003.
FOCMEC: FOcal MEChanism determinations,
{\sl International Handbook of Earthquake and Engineering
Seismology} (W. H. K. Lee, H. Kanamori, P. C. Jennings \& C.
Kisslinger, Eds.), Academic Press, San Diego, Chapter 85.12,
pp.~1629-1630.

\reference
Stein, R. S., 1999.
The role of stress transfer in earthquake occurrence,
{\sl Nature}, {\bf 402}, 605-609.

\reference
Storchak, D. A., D. Di Giacomo, I. Bondar, E. R. Engdahl, J.
Harris, W. H. K. Lee, A. Villaseqor \& P. Bormann (2013).
Public Release of the ISC-GEM Global Instrumental Earthquake
Catalogue (1900--2009),
{\sl Seismol.\ Res.\ Lett.}, {\bf 84}(5), 810-815.

\reference
Toda, S. \& R. Stein, 2003.
Toggling of seismicity by the 1997 Kagoshima earthquake
couplet: A demonstration of time-dependent stress transfer,
{\sl J. Geophys.\ Res.}, {\bf 108}(B12), Art.\ No.\ 2567.

\reference
Turner, H.H., 1914.
On a method of solving spherical triangles, and performing
other astronomical computations, by use of a simple table of
squares,
{\sl Monthly Notices Royal Astr. Soc.}, {\bf 75}(1), 530-541.

\reference
Wallace, T.,
A re-examination of the moment tensor solutions of the 1980
Mammoth Lakes earthquakes,
{\sl J. Geophys.\ Res.}, {\bf 90}, 11,171-11,176, 1985.

\reference
Wang, Q., D. D. Jackson \& Y. Y. Kagan, 2009.
California earthquakes, 1800-2007: a unified catalog with
moment magnitudes, uncertainties, and focal mechanisms,
{\sl Seism.\ Res.\ Lett.}, {\bf 80}(3), 446-457.

\reference
Wessel, P. \& W. H. F. Smith, 1998. 
New, improved version of Generic Mapping Tools, 
{\sl Eos Trans. AGU}, {\bf 79}, 579, doi:10.1029/98EO00426. 

\reference
Zechar, J.D., D. Schorlemmer, M.J. Werner, M.C. Gerstenberger,
D.A. Rhoades \& T.H. Jordan, 2013.
Regional Earthquake Likelihood Models I: First-Order Results,
{\sl Bull.\ Seismol.\ Soc.\ Amer.}, {\bf 103}(2a), 787-798.

\clearpage

\newpage

\renewcommand{\baselinestretch}{1.75}

\parindent=0mm

\newpage

\end{document}